 \documentclass[aps,prl,twocolumn,showpacs,superscriptaddress,longbibliography]{revtex4-2}
\usepackage{amsmath}
\usepackage{graphicx,epstopdf}
\usepackage{gensymb}
\newcommand{\be}{\begin{equation}}
\newcommand{\ee}{\end{equation}}
\newcommand{\bea}{\begin{eqnarray}}
\newcommand{\eea}{\end{eqnarray}}
\newcommand{\bse}{\begin{subequations}}
	\newcommand{\ese}{\end{subequations}}

\usepackage[utf8]{inputenc} % ok to keep for arXiv
\usepackage{xcolor}

\definecolor{darkred}{rgb}{0.7,0.0,0.0}
\definecolor{darkblue}{rgb}{0,0.02,0.45}
\definecolor{darkgreen}{rgb}{0.02,0.45,0.0}
\definecolor{violet}{rgb}{0.8,0.2,0.6}

\usepackage[colorlinks=true,bookmarks=false,
citecolor=darkblue,linkcolor=darkred,urlcolor=darkblue]{hyperref}

\begin{document}

\title{Anisotropic magnetoelastic coupling in the honeycomb magnet Na$_3$Co$_2$SbO$_6$}

\author{Prashanta K. Mukharjee}
\email{pkmukharjee92@gmail.com}
\affiliation{Experimental Physics VI, Center for Electronic Correlations and Magnetism, Institute of Physics, University of Augsburg, 86159 Augsburg, Germany}

\author{Sebastian Erdmann}
\affiliation{Experimental Physics VI, Center for Electronic Correlations and Magnetism, Institute of Physics, University of Augsburg, 86159 Augsburg, Germany}

\author{Lichen Wang}
\affiliation{Max-Planck-Institute for Solid State Research, Heisenbergstraße 1, 70569 Stuttgart, Germany}

\author{Julian Kaiser}
\affiliation{Experimental Physics VI, Center for Electronic Correlations and Magnetism, Institute of Physics, University of Augsburg, 86159 Augsburg, Germany}

\author{Anton Jesche}
\affiliation{Experimental Physics VI, Center for Electronic Correlations and Magnetism, Institute of Physics, University of Augsburg, 86159 Augsburg, Germany}

\author{Pascal Puphal}
\affiliation{Max-Planck-Institute for Solid State Research, Heisenbergstraße 1, 70569 Stuttgart, Germany}

\author{Masahiko Isobe}
\affiliation{Max-Planck-Institute for Solid State Research, Heisenbergstraße 1, 70569 Stuttgart, Germany}

\author{Matthias Hepting}
\affiliation{Max-Planck-Institute for Solid State Research, Heisenbergstraße 1, 70569 Stuttgart, Germany}

\author{Bernhard Keimer}
\affiliation{Max-Planck-Institute for Solid State Research, Heisenbergstraße 1, 70569 Stuttgart, Germany}

\author{Philipp Gegenwart}
\email{philipp.gegenwart@physik.uni-augsburg.de}
\affiliation{Experimental Physics VI, Center for Electronic Correlations and Magnetism, Institute of Physics, University of Augsburg, 86159 Augsburg, Germany}

\author{Alexander A. Tsirlin}
\email{altsirlin@gmail.com}
\affiliation{Felix Bloch Institute for Solid-State Physics, University of Leipzig, 04103 Leipzig, Germany}

\date{\today}

\begin{abstract}
	
We present magnetization and dilatometry measurements on the honeycomb cobaltate Na$_3$Co$_2$SbO$_6$ and map out its detailed field–temperature phase diagram down to sub-Kelvin temperatures. Our data for in-plane magnetic fields show a strongly anisotropic $c^*$-axis lattice response, which is dominated by the variation of Co--O--Co bond angles according to \textit{ab initio} calculations. At $T = 0.4$~K, the magnetization $M(B)$ exhibits step-like features that are also highly anisotropic. In the case of $B \parallel b$, a small hysteresis observed around the second field-induced magnetic transition ($B_{c2}$) indicates its first-order character, whereas divergence of the magnetic Gr\"uneisen parameter at $B_{c2}$ is suppressed upon cooling and signals the absence of quantum critical behavior upon entering the field-polarized state. None of our thermodynamic measurements provide evidence for a field-induced quantum spin liquid state near or above $B_{c2}$.
\end{abstract}

\maketitle

{\section{Introduction}}
The number of candidate quantum spin liquid (QSL) materials has been rapidly increasing in recent years. One important theoretical framework in this context is the Kitaev honeycomb model, which hosts an exactly solvable QSL ground state~\cite{Kitaev2}. The connection between the QSL state and the Kitaev model has been pivotal, driving intense research over the past decade. Honeycomb compounds based on $4d/5d$ transition metals, such as $\alpha$-RuCl$_3$ and (Li,Na)$_2$IrO$_3$, are leading examples, extensively studied as possible Kitaev QSL candidates~\cite{Winter2017}. Although these materials often show long-range antiferromagnetic (AFM) order due to additional interactions and deviations from ideal crystal symmetry, techniques like hydrostatic pressure~\cite{Shen2021}, doping~\cite{Kelly237203}, and applied magnetic fields have been explored to suppress this order and realize a Kitaev QSL state~\cite{Banerjee2018}. However, realizing Kitaev QSL in real materials is challenging because purely bond-directional Kitaev interactions are easily overwhelmed by unavoidable Heisenberg and off-diagonal terms, as well as interactions beyond nearest neighbors. 

Recently, $3d^7$ Co$^{2+}$ compounds, such as Na$_3$Co$_2$SbO$_6$, Na$_2$Co$_2$TeO$_6$, and BaCo$_2$(AsO$_4$)$_2$, have been proposed as new Kitaev candidates~\cite{Liu014407,Sano014408}. Theoretically, these honeycomb cobaltates are expected to host dominant Kitaev interactions because Co$^{2+}$ is a $J_{\mathrm{eff}} = \frac{1}{2}$ pseudospin ion, whereas the nearly $90^{\degree}$ \mbox{Co--O--Co} bonding geometry suppresses isotropic Heisenberg exchange and enhances bond-directional Kitaev terms. At the same time, they offer a practical advantage over $4d/5d$ systems, as they can be synthesized as large, high-quality single crystals with minimal structural disorder. These features render cobaltates an especially promising platform in comparison with their $4d/5d$ counterparts to study Kitaev-related physics. However, the non-Kitaev interactions are also significant or even dominant in cobaltates due to the subtle interplay of various hopping channels~\cite{winter2022,Liu054420}. 

Phenomenologically, the honeycomb cobaltates share several similarities with $\alpha$-RuCl$_3$ as a typical $4d/5d$ Kitaev material. All of them develop antiferromagnetic long-range order (LRO) in zero field, but this order is easily suppressed by applying magnetic field within the honeycomb plane, potentially leading to a field-induced Kitaev QSL. In the case of $\alpha$-RuCl$_3$, the character of the field-induced state above the critical field remains an active topic of debate~\cite{Kim2022}. Field-induced behavior of the honeycomb cobaltates is also under an active scrutiny~\cite{zhong2020,lee2025,Fang106701,ArnethL140402}. These materials often reveal distinct field-induced phases in the vicinity of the saturation field~\cite{MukharjeeL140407,Tu067304,hong2021,guang2023,zhang2023,pilch2023}, but magnetic order appears to persist within such phases~\cite{MukharjeeL140407,yao2023,Bera214419,shi2025}. An additional complexity arises from the multi-$q$ states reported in zero field in some of the cobaltates~\cite{gu2025}, in contrast to the simple zigzag state observed in $\alpha$-RuCl$_3$. 

\begin{figure}
	\includegraphics [width = \linewidth]{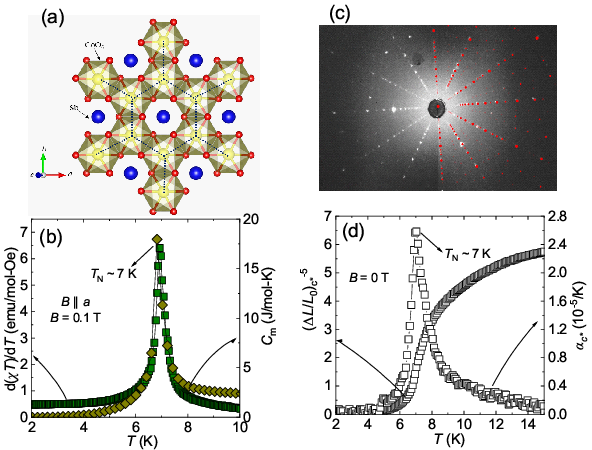}
	\caption{ (a) Crystal structure of Na$_3$Co$_2$SbO$_6$. The honeycomb network formed by CoO$_6$ octahedra spread in the $ab$-plane. (b) Fisher's heat capacity and magnetic specific heat vs. $T$ under zero magnetic field in the left and right $y$-axes respectively. (c) Left half: Reflections from the Laue pattern taken on a twin-free crystal, Right half: Simulation (red spots) of the Laue pattern using the \texttt{CRYSTAL MAKER} program. (d) Zero-field thermal expansion of NCSO.}
	\label{Fig1}
\end{figure}

Here, we focus on Na$_3$Co$_2$SbO$_6$ (NCSO) that crystallizes in the monoclinic space group $C2/m$ wherein edge-sharing CoO$_6$ octahedra form a honeycomb network in the $ab$ plane (see Fig. ~\ref{Fig1}(a))~\cite{Li041024}. Upon cooling in zero field, NCSO undergoes an AFM LRO below $T_N \simeq7$~K with a highly anisotropic field-induced response in the $ab$-plane~\cite{Li041024}. Magnetic ground state in zero field was long believed to be a simple zigzag phase~\cite{Viciu1060,wong2016,Yan2019}, but recent studies revealed a more intricate double-$q$ magnetic structure~\cite{Li041024,GuL060410} with residual spin dynamics~\cite{Miao134431}. The presence of such a multi-$q$ state, combined with pronounced in-plane anisotropy, render NCSO a compelling system for studying the effect of applied field on a honeycomb magnet.
 
Several recent works have explored field-induced effects in NCSO. At $T = 2$~K, in the ordered state, two distinct field-induced transitions are visible. The corresponding phase boundaries are anisotropic and appear at the critical fields $B_{c1} \simeq 1.25$ T and $B_{c2} \simeq 1.8$ T for $B \parallel a$ vs. $B_{c1} \simeq 0.7$ T and $B_{c2} \simeq 0.85$ T for $B \parallel b$ orientations~\cite{Li041024}. The region around $B_{c2}$ is of particular interest and has been studied with different probes such as nuclear magnetic resonance (NMR)~\cite{Hu054411}, THz spectroscopy~\cite{li2025}, heat transport~\cite{Fan085119}, ac-magnetostriction~\cite{Mi014417}, and magnetocaloric effect~\cite{Bestha2025v1}. However, the coupling between spins and lattice in NCSO is still largely unexplored. Magnetoelastic probes, in particular thermal expansion and magnetostriction, are very sensitive to phase transitions and magnetic anisotropy and can reveal signatures of quantum critical behavior or even a field-induced QSL state~\cite{Schonemann214432}.
 
These considerations raise three important questions for NCSO: (i) how strong is the magnetoelastic coupling, and how anisotropic is it with respect to the in-plane field direction? (ii) what is the nature of the field-induced transitions at $B_{c1}$ and $B_{c2}$? (iii) how do these transitions evolve toward $T=0$? Here, we address these questions by combining high-resolution dilatometry with other thermodynamic probes on high-quality twin-free single crystals of NCSO. We perform $c^*$-axis thermal expansion and magnetostriction measurements for magnetic fields applied along the in-plane $a$ and $b$ directions, and also report the magnetization and magnetic Gr\"uneisen parameter in the sub-Kelvin temperature range. We quantify the magnetoelastic response across all the field-induced changes and construct a detailed $B-T$ phase diagram. We find a pronounced in-plane anisotropy of the magnetoelastic coupling and clear signatures of the first-order behavior at $B_{c1}$. The apparent divergence of the magnetic Gr\"uneisen parameter near $B_{c2}$ is weak and does not increase at very low temperatures, thus giving evidence against a genuine field-induced quantum critical point in NCSO. Our results show that the intermediate-field regime of NCSO is governed by strongly anisotropic magnetic interactions and lacks any distinct signatures of a field-induced QSL state. \\

{\section{Methods}}
High-quality single crystals of NCSO were grown using an optimized flux method (see SM). The quality of these crystals was rigorously verified through extensive Laue mapping (Fig.~\ref{FigS1}) and thermodynamic measurements. The NCSO crystals often exhibit twinned structural domains. For this study, we selected a twin-free crystal measuring 1 × 1 × 0.1 mm$^3$. Laue analysis confirmed the absence of any structural domains in the chosen crystal. Notably, the twin-free crystal displayed only two field-induced features for each field direction (at 2\,K), whereas twinned crystals exhibited multiple transitions (not shown) for a given direction of the applied field. Furthermore, angle-dependent susceptibility measurements confirm the absence of twinning in the investigated crystal, as evidenced by the clear two-fold symmetry observed (see Fig.~\ref{FigS8}).

\begin{figure*}[htbp]
	\centering
	\includegraphics[width = 0.8\textwidth]{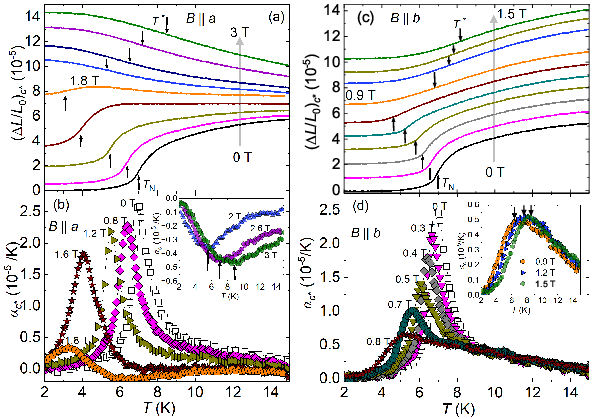}
	\caption{\label{Fig2} (a, c) Temperature dependence of the relative length changes for $B\parallel a$ and $B \parallel b$, respectively (warming curves are shown). The upward arrows indicate the position of the $T_N$. The downward arrows mark the position of the crossover temperature $T^*$ above $B_{c2}$. (b, d) The corresponding plots of the linear TE coefficient, $\alpha_{c^*}$ vs. $T$  for $B\parallel a$ and $B \parallel b$, respectively. The insets show the data measured above $B_{c2}$. The arrows indicate the position of the broad maximum/minimum in $\alpha_{c^*}$.}
\end{figure*}

Magnetization measurements were performed using a Quantum Design Magnetic Property Measurement System (QD-MPMS3). For measurements conducted below 2\,K, a He$_3$ attachment was utilized. Magnetic isotherms were collected at a sweep rate of 200 Oe/sec in stable mode. Angle-dependent magnetization was recorded by employing the sample rotation stage of the MPMS3. The specific heat was measured using a standard relaxation technique within the QD Physical Property Measurement System (PPMS). The magnetic Gr\"uneisen parameter is estimated from the magnetocaloric effect using alternating-field technique~\cite{Tokiwa013905}. 

Thermal expansion ($\Delta L/L_0$) and magnetostriction ($\lambda$) were measured along $c^*$-axis [$c^* = \frac{1}{c \sin\beta}$] using a high-resolution capacitance dilatometer following the procedure described in Ref.~\cite{Kuchler083903}. The temperature and field scans of the length change were carried out at rates of 0.3 K/min and 20 Oe/sec, respectively. Our dc-measurements directly probe the out-of-plane expansion of NCSO, in contrast to Ref.~\cite{Mi014417} where in-plane lattice expansion was measured in the ac-mode, resulting in the field derivative of the magnetostriction ($d\lambda/dB$) and not the magnetostriction itself.

Effect of magnetic order on the lattice parameters and atomic structure of NCSO was studied by density-functional theory (DFT) band-structure calculations performed in the \texttt{VASP} code~\cite{vasp1,vasp2} using the Perdew-Burke-Ernzerhof version of the exchange-correlation potential~\cite{pbe96}. All calculations included spin-orbit (SO) coupling and the mean-field correction for correlation effects in the Co $3d$ shell, with the on-site Coulomb repulsion $U_d=5$\,eV and Hund's coupling $J_d=1$\,eV~\cite{szaller2025}. \\

\section{Results}

{\subsection{Thermal Expansion}}

Figure~\ref{Fig2} displays the thermal expansion (TE) data for NCSO up to 15\,K, with the background contribution subtracted. The background contribution was determined by measuring length changes in a piece of copper of the same dimensions as the NCSO crystal (see Fig.~\ref{FigS4}b). In zero magnetic field, the relative length change $(\Delta L/L_0)_c$ (with $L_0$ defined as the $c^*$-axis sample length at $T = 300$~K and 0\,T) decreases upon cooling, showing an S-shaped anomaly around the Néel temperature $T_N \sim 7$~K, as seen in Fig.~\ref{Fig1}d, indicating that NCSO contracts along the $c^*$-axis upon entering the magnetically ordered state, in contrast to $\alpha$-RuCl$_3$ where cooling across $T_N$ leads to the sample expansion~\cite{Gass2020}.

\begin{figure*}
	\includegraphics[width = 0.8\textwidth]{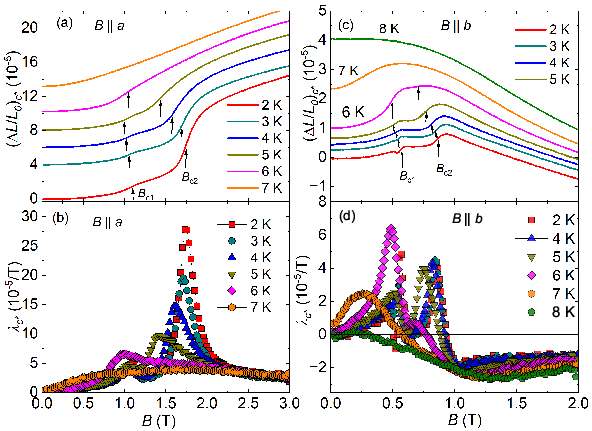}
	\caption{\label{Fig3} (a,c) Field dependence of relative length changes at constant temperature for $B\parallel a$ and $B \parallel b$, respectively (the data are measured upon increasing the field). The curves are shifted vertically for clarity. The arrows indicate the positions of $B_{c1}$ and $B_{c2}$. (b,d) Their corresponding coefficients of magnetostriction vs. field.}
\end{figure*}

The cooling and warming curves show a perfect reversibility around $T_N$, suggesting the second-order nature of this transition (see Fig.~\ref{FigS10}). The corresponding linear thermal expansion coefficient, $\alpha = \frac{1}{L_0} \frac{d\Delta L}{dT}$, shown on the right $y$-axis of Fig.~\ref{Fig1}d (right $y$-axis), reveals a pronounced positive peak that signals the onset of the AFM order. The substantial lattice changes [$(\Delta L/L_0)_{c^*} \sim$ 6 $\times 10^{-5}$] observed around $T_N$ suggest the presence of magnetoelastic coupling in NCSO. For comparison, we have also included the zero-field relative length changes of various honeycomb magnets in Table~\ref*{Tab1}.

\begin{figure*}[htbp]
	\centering
	\includegraphics{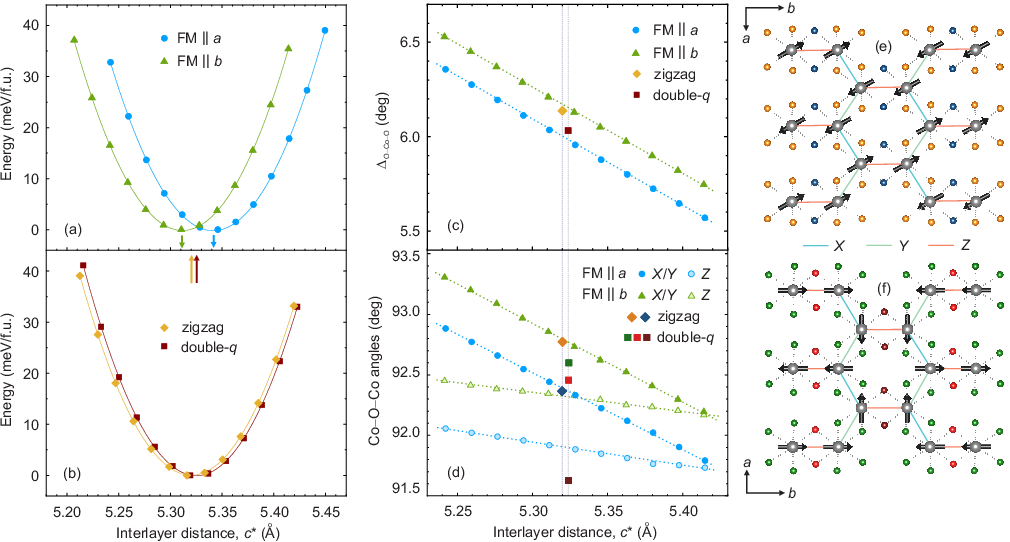}
	\caption{\label{Fig4} (a,b) Total energy depending on the interlayer distance $c^*$ for different spin configurations of NCSO. The positions of the energy minima are indicated by arrows. (c) Deformation of the CoO$_6$ octahedra gauged by $\Delta_{\rm O-Co-O}$, the average deviation of O--Co--O angles from the median value of $90.0^{\circ}$. (d) Co--O--Co bond angles. The color codes for the zigzag and double-$q$ states correspond to the different oxygen atoms, as depicted in panels (e) and (f), respectively. Only the bond angles at the position of the energy minimum are shown in each case.
 }
\end{figure*}

Next, we study the effect of an applied magnetic field on the AFM order for both in-plane field directions (see Fig.~\ref{Fig2}). At low fields, the length changes are similar to those in zero field, but as the magnetic field approaches $B_{c2}$, the responses in the two orientations diverge significantly. For $B \parallel a$, the S-shaped anomaly in $(\Delta L/L_0)_{c^*}$ at $T_N$ evolves into a broad minimum around the critical field $B_{c2} \sim 1.8$~T, and at higher fields the relative length change increases upon cooling, indicating a lattice expansion. In contrast, for $B \parallel b$, $(\Delta L/L_0)_{c^*}$ continues to decrease smoothly toward low temperatures for fields above $B_{c2} \sim 0.85$~T, indicating that the $c^*$-axis length keeps shrinking for this field configuration. To track these field-induced anomalies, we plot the corresponding values of $\alpha_{c^*}$ in Fig.~\ref{Fig2}b and Fig.~\ref{Fig2}d for $B\parallel a$ and $B \parallel b$, respectively. With increase in the magnetic field, the sharp peak in $\alpha_{c^*}$ shifts to lower temperatures, reflecting the suppression of AFM order, consistent with the trends observed in DC magnetic susceptibility and specific heat (Fig.~\ref{FigS2} and Fig.~\ref{FigS5}). Near $B_{c2}$, the peak of $\alpha_{c^*}$ broadens significantly, and a distinct behavior emerges beyond $B_{c2}$: while a broad maximum is seen for both field orientations, the absolute values of $\alpha_{c^*}$ differ notably. For $B \parallel b$, $\alpha_{c^*}$ shows a broad maximum with the positive values, but for $B\parallel a$, $\alpha_{c^*}$ becomes negative, and eventually a minimum in $\alpha_c$ is observed. Similar type of a sign change in $\alpha_{c}$ has been also reported in Na$_2$Co$_2$TeO$_6$ ~\cite{ArnethL140402}. The broad maximum/minimum in $\alpha_{c^*}$ observed above $B_{c2}$ marks a crossover from the field-induced saturated state into the paramagnetic state. The contrasting behaviors observed for the two in-plane directions highlight the strong in-plane anisotropy of the magnetoelastic coupling in NCSO. The temperature values of the sharp anomalies observed in the ordered state and the positions of the broad maxima/minima in the field-polarized state are compiled in the $B-T$ phase diagram shown in Fig.~\ref{Fig7}.\\

{\subsection{Magnetostriction}}

To monitor the field-induced changes in NCSO, we measured magnetostriction (MS) at constant temperature over the temperature range 2\,K $\leq T \leq$ 8\,K (Fig.~\ref{Fig3}). For $T \leq 5$~K, the $c^*$-axis length increases with increasing magnetic field for both in-plane field directions, as evidenced by the monotonic rise in $(\Delta L/L_0)_{c^*}$. Both the transitions at $B_{c1}$ and $B_{c2}$ appear as kinks in the length change data. In the high-temperature regime above 6\,K (7\,K), the length change becomes featureless for $B\parallel a$ ($B \parallel b$). Similar to the length change measured at constant field, the field-dependent length changes also show contrasting behaviors around $B_{c2}$ for the $a$ and $b$-directions. While $(\Delta L/L_0)_{c^*}$ continues to increase beyond 1.8\,T for $B \parallel a$, it shrinks for $B \parallel b$ at all measured temperatures. Notably, the length change for $B \parallel a$ is nearly ten times larger than that for $B \parallel b$, indicating a more significant magnetoelastic coupling when field is applied along $a$.

The corresponding coefficients of the linear magnetostriction ($\lambda = \frac{1}{L_0} \frac{d\Delta L}{dB}$), are plotted in Fig.~\ref{Fig3}b and Fig.~\ref{Fig3}d for both in-plane directions. At temperatures below 5\,K, two sharp anomalies at $B_{c1}$ and $B_{c2}$, are clearly evident as distinct peaks in the $\lambda_{c^*}$ versus $B$ curves, whereas no additional anomaly is observed below $B_{c1}$. Such a low-field anomaly has been seen in various field-dependent thermodynamic measurements on $\alpha$-RuCl$_3$~\cite{Gass2020,Bachus097203,Schonemann214432} where it is usually interpreted as the domain repopulation within the zigzag state. The absence of this anomaly in NCSO corroborates the monodomain double-$q$ ground state in this material.

The field hysteresis at $B_{c1}$ (see Fig.~\ref{FigS6} and see Fig.~\ref{FigS7}) suggests the first-order nature of this transition. The hysteresis persists up to 4\,K (5\,K) the for $a$ ($b$)-directions, and disappears upon approaching $T_N$. Conversely, the up $\leftrightarrow$ down data at $B_{c2}$ are reversible at all measured temperatures. Above 5\,K (6\,K) with $B\parallel a$ ($B \parallel b$), the two transitions merge, whereas at even higher temperatures $\lambda(B)$ becomes featureless. The corresponding critical fields for each temperature are consistent with the results obtained from the $M(B)$ measurements, as shown in Fig.~\ref{Fig7}. We also find a close similarity between field dependences of $\lambda$ and $dM/dB$, as shown in Figs.~\ref{FigS6} and~\ref{FigS7}].\\

{\subsection{Effect of out-of-plane strain}}

We now explore the origin of anisotropic magnetostriction using DFT+$U$+SO calculations. To this end, we relaxed atomic positions and lattice parameters of NCSO for fixed values of $c^*$ assuming different spin configurations: the ferromagnetic state with the spins pointing along either $a$ or $b$, as well as the zigzag and double-$q$ states as possible magnetic structures in zero field. Fig.~\ref{Fig4}(a,b) displays the parabolic dependence of total energy on $c^*$, as expected for the elastic strain applied to the material. The slope of these curves returns an effective elastic constant of 230\,GPa, which is within the expected range for an ionic compound.

Remarkably, different spin configurations lead to different equilibrium values of $c^*$. Their hierarchy follows our experimental results, namely, the interlayer distance in either the zigzag or double-$q$ state is shorter compared to the FM state for $B\parallel a$ and longer than in the FM state for $B\parallel b$. The relative changes in $c^*$ are somewhat larger than the values of $(\Delta L/L_0)_{c^*}\simeq 1\times 10^{-4}$ observed experimentally. This is probably an artifact of DFT+$U$+SO where Co$^{2+}$ is treated as a spin-$\frac32$ ion and not as the $j_{\rm eff}=\frac12$ ion, leading to overestimated magnetoelastic energies, with somewhat exaggerated changes in $c^*$ between the different spin configurations. Importantly, though, the trends in $c^*$ are reproduced, and the effect of spin configurations on individual structural parameters can be explored in detail. 

The difference in $c^*$ between the FM states polarized along $a$ and $b$ could be related either to single-ion physics of Co$^{2+}$, or to anisotropic magnetic couplings. These effects are controlled by the local geometry of the CoO$_6$ octahedra and the Co--O--Co bond angles, respectively. The monoclinic crystal structure of NCSO gives rise to a rather complex deformation of the CoO$_6$ octahedra, but the median O--Co--O angle remains very close to $90.0^{\circ}$ for all strains. Therefore, we quantify the octahedral deformation by the average absolute deviation of the O--Co--O angles from this median value. Fig.~\ref{Fig4}(c) shows that the tensile strain along $c^*$ renders the octahedra more symmetric, but for a given $c^*$, the difference in the deformation between the FM states polarized along $a$ and $b$ is about $0.15^{\circ}$ only. 

The Co--O--Co bond angles in Fig.~\ref{Fig4}(d) show a much larger difference of about $0.4^{\circ}$ depending on the spin direction. This effect is a fingerprint of anisotropic exchange interactions, because isotropic Heisenberg terms are insensitive to the direction of spin. The presence of anisotropic exchange is in agreement with the earlier findings of either Kitaev~\cite{songvilay2020,sanders2022} or off-diagonal anisotropy~\cite{Bestha2025v1} terms being dominant in NCSO. 

The monoclinic structure of NCSO features two distinct Co--O--Co bond angles that corresponds to the $X/Y$- and $Z$-bonds in the Kitaev frame. The Co--O--Co angles for the Z-bonds are systematically lower than those for the $X/Y$-bonds. While both angles decrease with increasing $c^*$, they change at a different rate and approach each other at about 2\% of the tensile strain. It is further instructive to compare these angles with the optimized values obtained for the zigzag and double-$q$ states [Fig.~\ref{Fig4}(d)]. The zigzag state involves FM spin-spin correlations on the $Z$-bond and one of the $X,Y$-bonds. From Goodenough-Kanamori-Anderson rules, one expects that lower Co--O--Co angles should favor a FM coupling. Indeed, the respective angle is about $92.4^{\circ}$ for the $Z$-bond, but the second FM correlation has to set in on the $Y$-bond with the larger angle. The double-$q$ state is accompanied by a much stronger structural deformation. The FM correlations are now confined to the $Z$-bonds, and half of them -- those with the spins along $a$ -- develop much smaller bond angles of $91.6^{\circ}$, which are comparable to the $Z$-bond angles in the FM state polarized along $a$. Altogether, one sees consistent changes in the Co--O--Co bond angles with respect to both spin direction and the type of magnetic order on each bond. \\

{\subsection{Magnetic Gr\"uneisen parameter}}

\begin{figure}
	\includegraphics[width = \linewidth]{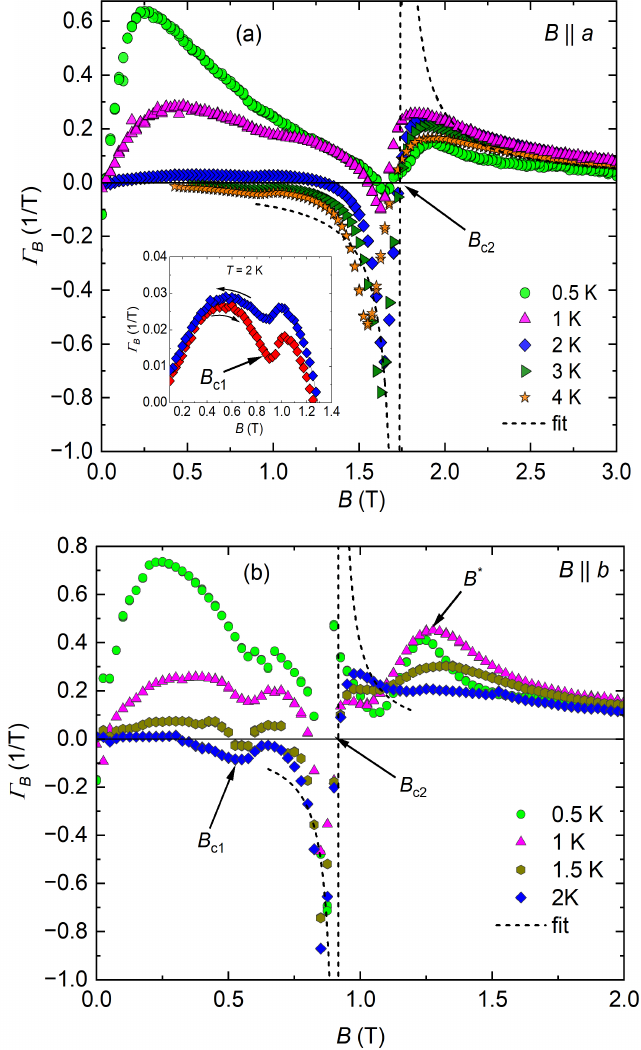}
	\caption{\label{Fig5} Magnetic field dependence of $\Gamma_B$ for (a) $B\parallel a$ and (b) $B\parallel b$. The inset in (a) shows the hysteresis around $B_{c1}$. The dashed lines are the fit as described in the text.}
\end{figure}

We now return to the field-induced phase transitions in NCSO and study them in more detail using magnetocaloric effect (MCE) measured below the N\'eel temperature (Figure~\ref{Fig5}). The magnetic Gr\"uneisen parameter [$\Gamma_B=(1/T)\,(\partial T/\partial B)_S$] which equals the relative adiabatic magnetocaloric effect displays anomalies at $B_{c1}$ and $B_{c2}$. For both field directions, the anomaly at $B_{c1}$ appears as a broad minimum, which is more pronounced for $B\parallel b$. Upon approaching $B_{c2}$, the shape of $\Gamma_B$ changes abruptly and shows a divergent behavior with a sign change. The up-down sweep performed for $B \parallel a$ reveals a pronounced field hysteresis around $B_{c1}$, but essentially no hysteresis near $B_{c2}$, in agreement with our magnetostriction and magnetization data.

For $B \parallel a$, several experimental results point to unusual behavior in the vicinity of $B_{c2}$. The NMR spin-lattice relaxation rate shows an enhancement of low-energy spin fluctuations together with a gapped regime around $B_{c2}$~\cite{Hu054411}. Thermal transport measurements likewise report a pronounced minimum in the thermal conductivity $\kappa$ at $B_{c2}$, attributed to an enhanced phonon scattering by magnetic fluctuations~\cite{Fan085119}. In addition, ac-magnetostriction experiments identify a tricritical point close to $B_{c2}$~\cite{Mi014417}. While these observations raise the possibility of the field-induced quantum criticality, they do not establish it conclusively. In our data, the field dependence of $\Gamma_B(B)$ (Fig.~\ref{Fig5}) shows a divergence near $B_{c2}$, reminiscent of field-induced quantum critical points in itinerant magnets~\cite{Tokiwa226402,Tokiwa116401}. The sign change of $\alpha_{c^*}$ at $B_{c2}$ (Fig.~\ref{Fig2}b) is also consistent with such a scenario, although it is not a direct proof of quantum criticality. 

To assess whether a quantum-critical description is applicable, we therefore fitted the field dependence of $\Gamma_B$ near $B_{c2}$ using a power law of the form $\Gamma_B = G/(B - B_c)$, where $G$ is a constant~\cite{Garst205129}. In these fits, $B_{c2}$ was fixed to the zero crossing values in $\Gamma_B$. The dashed lines in Fig.~\ref{Fig5} show the corresponding fits, yielding the prefactors $G$ about 0.07 and 0.03 for $B \parallel a$ and $B \parallel b$, respectively. These prefactors are much smaller compared to systems with established field-induced quantum criticality~\cite{Tokiwa226402,Tokiwa116401}. In addition, the apparent divergence of $\Gamma_B$ becomes weaker on cooling, whereas for a genuine quantum critical point it should be enhanced at low temperatures. Taken together, the small prefactors and the temperature evolution of the $\Gamma_B$ divergence suggest the absence of quantum critical behavior in NCSO at $B_{c2}$.

Extending thermodynamic measurements on NCSO below 2\,K not only evaluates quantum criticality in NCSO, but also exposes some further unusual features of this material. Fig.~\ref{Fig5} shows that $\Gamma_B$ is small below $B_{c1}$ at 2\,K, but at lower temperatures it increases significantly and develops a broad feature around 0.3\,T. Since this feature is not observed at higher temperatures, its origin is probably unrelated to domain repopulation. The monodomain multi-$q$ ground state of NCSO further renders such a scenario unlikely. In order to shed more light on the behavior of NCSO at sub-Kelvin temperatures, we turn to the magnetization measurements below 2\,K.
\\

{\subsection{Magnetization below 2\,K}}

\begin{figure}
	\includegraphics[width = \linewidth]{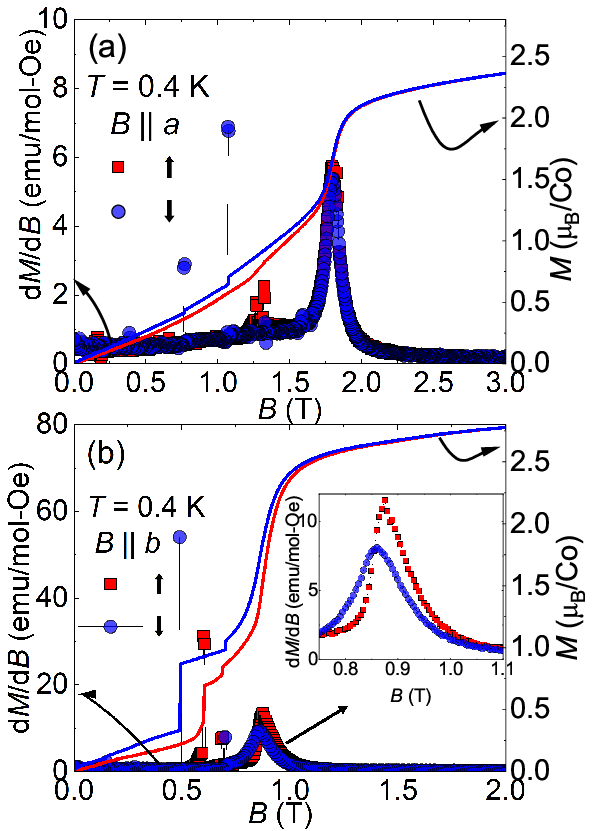}
	\caption{\label{Fig6} Magnetization isotherms measured at $T$ = 0.4\,K for (a) $B \parallel a$ and (b) $B \parallel b$ directions. The inset in (b) shows the tiny hysteresis observed around $B_{c2}$.}
\end{figure}

Figure~\ref{Fig6} shows the field-dependent magnetization $M(B)$ of NCSO at $T = 0.4$~K (the data for other temperatures are shown in Fig.~\ref{FigS3}). As seen in Fig~\ref{Fig6} the magnetization undergoes abrupt changes with the field. It exhibits several clear steps that are especially pronounced for $B \parallel b$. Though the number of steps is relatively small, each step is well-defined (sharp peaks in $dM/dB$), thus pointing to abrupt changes of the spin configuration rather than a gradual canting of the magnetic moments.

Similar step-like behavior has been reported in other Co-based magnets. In BaCo$_2$(AsO$_4$)$_2$, field-dependent magnetization measurements at $T$ = 0.14\,K reveal a series of small steps~\cite{Devillez2025}. These have been interpreted in terms of slow domain and defect dynamics and avalanche-like jumps between metastable states of a frustrated magnet. A very similar behavior is also found in BaCo$_2$(PO$_4$)$_2$~\cite{Wang2023}, where magnetostriction below 1\,K shows jumps and tiny plateaus. Our data for NCSO, although showing fewer and more widely spaced steps, point in the same general direction. 

Two further features are worth mentioning. First, the $B_{c1}$ transition at 0.4\,K is accompanied by a large and abrupt change in the magnetization for $B \parallel b$ but not for $B \parallel a$. This difference reinforces the idea of Ref.~\cite{Bestha2025v1} on the different nature of the field-induced state between $B_{c1}$ and $B_{c2}$ for the two field directions, $\frac13$-AFM for $B \parallel a$ vs. canted zigzag for $B \parallel b$, although the relation of these states to the size of the magnetization steps remains to be established. Second, all measurements above 2\,K consistently show second-order nature of the transition at $B_{c2}$, but a small hysteresis appears in the $M(B)$ data for $B \parallel b$ (see inset of Fig.~\ref{Fig6}b). This feature indicates that the second-order transition at $B_{c2}$ becomes first-order at sub-Kelvin temperatures and further excludes the presence of a quantum critical point at $B_{c2}$. \\

{\section{Phase Diagram and Discussion}}
\begin{figure}
	\includegraphics[width = \linewidth]{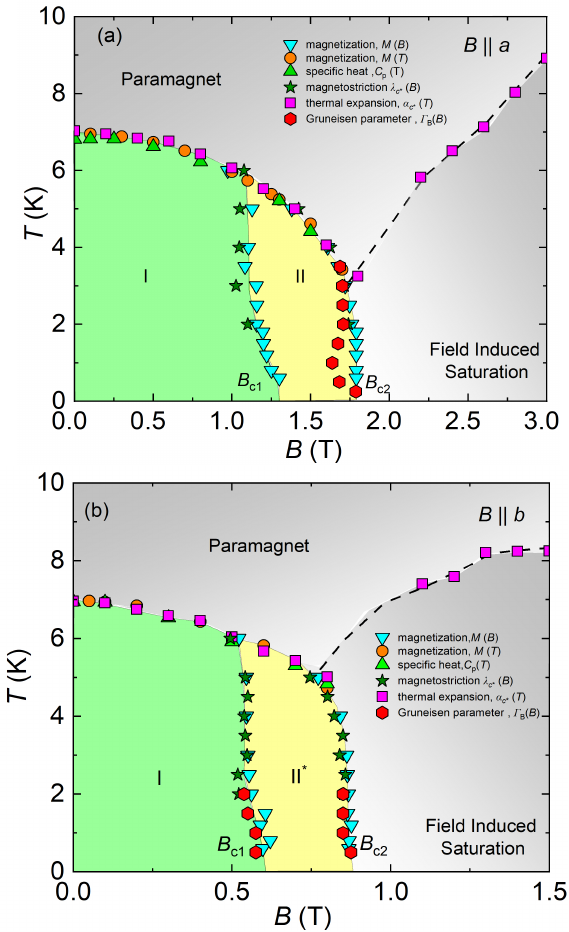}
	\caption{\label{Fig7} Field-temperature phase diagram of NCSO constructed from various thermodynamic measurements for (a) $B \parallel a$ (b) $B \parallel b$ directions. The anomalies in the $\Gamma_B$ points in panel (a) may be due to the first-order nature of the transition at $B_{c2}$ below 2\,K.}
\end{figure}

Figure~\ref{Fig7} shows the field–temperature phase diagram of NCSO, constructed from all our measurements discussed in the previous section. The phase boundaries are determined from the anomalies in the magnetization, specific heat, and dilatometry, namely, from the peaks in $d(\chi T)/dT$, $dM(B)/dB$, $\alpha_{c^*}(T)$, $\lambda_{c^*}(T)$, and $C_p(T)$. For the MCE data, we take $B_{c1}$ as the position of the broad minimum in $\Gamma_B$, and identify $B_{c2}$ by the zero crossings of $\Gamma_B$. Above $B_{c2}$, $\alpha_{c^*}(T)$ exhibits a broad maximum for both in-plane field directions; the position of this maximum at each field is used to mark the crossover between the high-temperature paramagnetic regime and the field-induced saturated state. Two magnetically ordered states are identified for each field direction. Phase I entails the double-$q$ order according to neutron diffraction~\cite{GuL060410}, whereas phase II is described as the AFM-1/3 state for $B \parallel a$~\cite{Li041024} and canted zigzag state for $B \parallel b$~\cite{Bestha2025v1}. Our phase boundaries agree well with those reported previously above 2\,K. They extend toward sub-Kelvin temperatures almost vertically, the main difference being the change in the nature of the transition from second-order toward first-order at $B_{c2}$.

The initial pressure dependence of the magnetic transitions can be inferred from the thermal expansion and magnetostriction data. The zero-field transition at $T_N \simeq 7$~K is second-order (see Fig.~\ref*{FigS10}). For such a transition, the uniaxial pressure dependence of $T_N$ along the $c^*$-axis can be estimated using the Ehrenfest relation $dT_N/dp_{c^*} = V_m T_N \Delta \alpha_{c^*} / \Delta C_p$, where $V_m = 8.04 \times 10^{-5}\,\mathrm{m}^3/\mathrm{mol}$ is the molar volume, $\Delta \alpha_{c^*}$ is the jump in the linear TE coefficient at $T_N$, and $\Delta C_p$ is the specific-heat change at the transition. Using $T_N \simeq 7\,\mathrm{K}$, $\Delta \alpha_{c^*} = 2.5 \times 10^{-5}\,\mathrm{K}^{-1}$, and $\Delta C_p = 18\,\mathrm{J/mol\,K}$, we obtain $dT_N/dp_{c^*} \simeq 0.77\,\mathrm{K/GPa}$, indicating that $T_N$ should increase if pressure is applied along the $c^*$-axis. Further insight into the pressure response follows from the Maxwell relation $\lambda V = - (dM/dp)_{T,B}$, where $\lambda$ is the magnetostriction coefficient, $V$ is the sample volume, and $M$ is the magnetization. Although our measurements are limited to linear magnetostriction only along the $c$-axis, rather than the full volume magnetostriction, the positive $\lambda_{c^*}$ implies $(dM/dp)_{T,B} < 0$. Hence, the magnetization is expected to decrease upon compression along the $c^*$-axis of NCSO.

Now, we discuss the region close to $B_{c2}$, which is of particular interest in the context of possible field-induced QSL. The behavior of $\alpha$-RuCl$_3$ in this region is still vividly debated~\cite{Kim2022}. In the case of BaCo$_2$(AsO$_4$)$_2$, the formation of a gapless QSL has been proposed on the basis of thermal conductivity measurements~\cite{Tu067304}, but the observation of persistent magnetic Bragg peaks in this region~\cite{MukharjeeL140407} argues against this scenario. A similar discussion exists for Na$_2$Co$_2$TeO$_6$, where it is also debated whether a QSL phase appears above $B_{c2}$~\cite{Fang106701}. In NCSO, the in-plane anisotropy renders the behavior above $B_{c2}$ strongly dependent on the field direction. 

For $B \parallel a$, NMR measurements indicated a possible additional phase extending from $B_{c2}$ up to 3.0\,T according to the nonmonotonic evolution of the spin gap~\cite{Hu054411}, but our data clearly exclude any phase transitions beyond $B_{c2}$. A more plausible scenario is the crossover from second-order toward first-order transition upon cooling and phase coexistence around $B_{c2}$, which would also be consistent with the observation of magnetic Bragg peaks also above $B_{c2}$~\cite{Li041024}. 

The $B \parallel b$ case is more intriguing, because $\Gamma_B$ shows a peak at $1.2-1.3$\,T, well above $B_{c2}$ (Fig.~\ref{Fig5}). This feature is reminiscent of the shoulder in $\Gamma_B$ reported in $\alpha$-RuCl$_3$ in the field range of the putative QSL~\cite{bachus2021}, but that shoulder can be understood in terms of a level crossing of two low-lying excited states and does not signal a QSL~\cite{Bachus097203}. A similar explanation may apply to NCSO too. Ref.~\cite{Bestha2025v1} further reports a shoulder-like anomaly in the specific heat for $B \parallel b$, which was discussed in terms of short-range magnetic order. Another important observation is that the shoulder in $\Gamma_B$ appears below 2\,K only, concurrent with the peak in $\Gamma_B$ around 0.3\,T and the steps in the magnetization. All these features may be potentially related to slow dynamics, possibly of domain boundaries, and clearly deserve further investigation. 
\\

{\section{Summary}}

In summary, we have investigated the thermodynamic properties of the honeycomb magnet Na$_3$Co$_2$SbO$_6$ using high-resolution dilatometry, magnetocaloric-effect, magnetization, and specific-heat measurements and extended the phase diagram of this material into the sub-Kelvin temperature range. In the ordered state, two field-induced transitions at $B_{c1}$ and $B_{c2}$ are observed in all probes. Whereas the former transition is first-order, the latter transition is second-order above 2\,K and gradually becomes first-order at lower temperatures. Anisotropic magnetoelastic coupling is reproduced using \textit{ab initio} calculations and interpreted as an effect of changes in the Co--O--Co bond angles caused by the out-of-plane strain. Our data rule out quantum critical behavior near $B_{c2}$ and further show intriguing step-like features in the magnetization, which are  characteristic of metastable states. There is no thermodynamic evidence for a field-induced QSL state in Na$_3$Co$_2$SbO$_6$.\\

{\section{acknowledgements}}

This work was funded by the Deutsche Forschungsgemeinschaft (DFG, German Research Foundation) -- TRR 360 -- 492547816 (subproject B1 and B2). The authors gratefully acknowledge the use of computing resources of the ALCC HPC cluster (Institute of Physics, University of Augsburg).\\

The data supporting the findings of this study are available \cite{NCSO_data}.\\

\noindent \textcolor{red}{$^*$}pkmukharjee92@gmail.com\\
\textcolor{red}{$^\dagger$}philipp.gegenwart@physik.uni-augsburg.de\\
\textcolor{red}{$^\ddag$}altsirlin@gmail.com
\bibliography{reff_NCSO}
\widetext
\clearpage
%\end{document}
\begin{center}
	\textbf{\large Supplementary Material for \\
	Anisotropic magnetoelastic coupling in Na$_3$Co$_2$SbO$_6$}
\end{center}
\author{Prashanta K. Mukharjee}
\email{pkmukharjee92@gmail.com}
\affiliation{Experimental Physics VI, Center for Electronic Correlations and Magnetism, Institute of Physics, University of Augsburg, 86159 Augsburg, Germany}

\author{Lichen Wang}
\affiliation{Max-Planck-Institute for Solid State Research, Heisenbergstraße 1, 70569 Stuttgart, Germany}

\author{Anton Jesche}
\affiliation{Experimental Physics VI, Center for Electronic Correlations and Magnetism, Institute of Physics, University of Augsburg, 86159 Augsburg, Germany}

\author{Julian Kaiser}
\affiliation{Experimental Physics VI, Center for Electronic Correlations and Magnetism, Institute of Physics, University of Augsburg, 86159 Augsburg, Germany}

\author{Matthias Hepting}
\affiliation{Max-Planck-Institute for Solid State Research, Heisenbergstraße 1, 70569 Stuttgart, Germany}

\author{Philipp Gegenwart}
\email{philipp.gegenwart@physik.uni-augsburg.de}
\affiliation{Experimental Physics VI, Center for Electronic Correlations and Magnetism, Institute of Physics, University of Augsburg, 86159 Augsburg, Germany}

\author{Alexander A. Tsirlin}
\email{altsirlin@gmail.com}
\affiliation{Felix Bloch Institute for Solid-State Physics, University of Leipzig, 04103 Leipzig, Germany}

\setcounter{equation}{0}
\setcounter{figure}{0}
\setcounter{table}{0}
\setcounter{page}{1}
\makeatletter
\setcounter{section}{0}
\renewcommand{\thesection}{S-\Roman{section}}
\renewcommand{\thetable}{S\arabic{table}}
\renewcommand{\theequation}{S\arabic{equation}}
\renewcommand{\thefigure}{S\arabic{figure}}
\renewcommand{\bibnumfmt}[1]{[S#1]}
\renewcommand{\citenumfont}[1]{S#1}

{\subsection{Crystal Growth}}
Single crystals of NCSO were grown using a flux method adapted from  Ref.~\cite{Yan2019} with carefully optimized growth parameters. High-quality  
and dark ruby-red crystals were obtained. Our samples exhibit a higher $T_N$ compared to that reported in Ref.~\cite{Yan2019}, while the overall fundamental characteristics are consistent with those described in Ref.~\cite{Li041024}.

{\subsection{Specific heat}}

Due to the very thin dimensions of the investigated crystal, a vertical puck setup from Quantum Design was used to ensure precise alignment of the crystal in the in-plane orientations. To subtract the lattice contribution to the specific heat (see Fig.~\ref{FigS4}), a non-magnetic compound, Na$_3$Zn$_2$SbO$_6$, was synthesized. To account for the different formula masses of Na$_3$Zn$_2$SbO$_6$ and Na$_3$Co$_2$SbO$_6$, we followed a scaling procedure as detailed in~\cite{Anand184403}.
\clearpage

\begin{figure}[htb]
	\includegraphics[width= 14 cm]{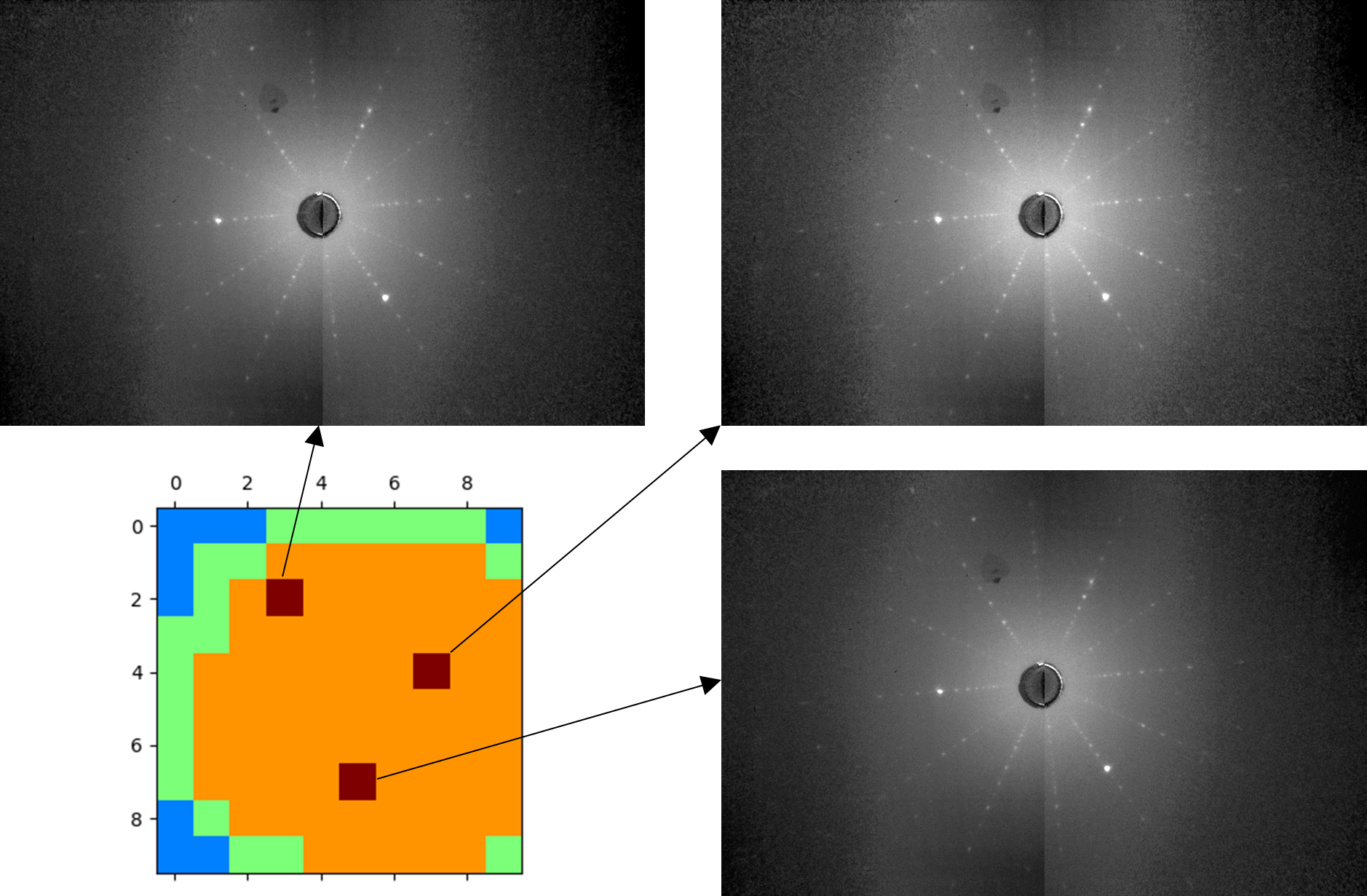}
	\caption{A Laue back-scattering grid analysis was used to identify multiple domains in the NCSO single crystal. In the color-coded diffraction images, orange indicates clear diffraction patterns, green represents weaker patterns (from the crystal edges), and blue shows areas with no patterns (corresponding to the sample holder). The consistent crystallographic orientation across various points on the crystal surface confirmed the absence of multiple growth domains.
		\label{FigS1}} 
\end{figure}

\begin{figure}[htb]
	\includegraphics[width= 17 cm]{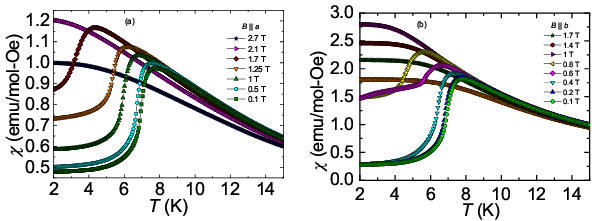}
	\caption{Temperature dependent DC magnetic susceptibility for (a) $B \parallel a$ (b) $B \parallel b$ orientations.
		\label{FigS2}}
\end{figure}

\begin{figure}
	\includegraphics[width= 17 cm]{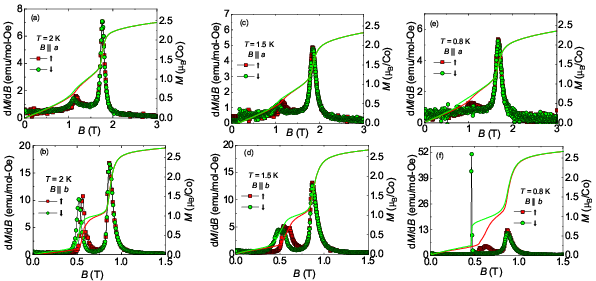}
	\caption{Field-dependent magnetization (right axes) and its derivative $\mathrm{d}M/\mathrm{d}B$ (left axes) obtained in increasing and decreasing sweeps at $0.8$–$2$ K for $B \parallel a$ and $B \parallel b$.
		\label{FigS3}}
\end{figure}

\begin{figure}
	\includegraphics[width= 17 cm]{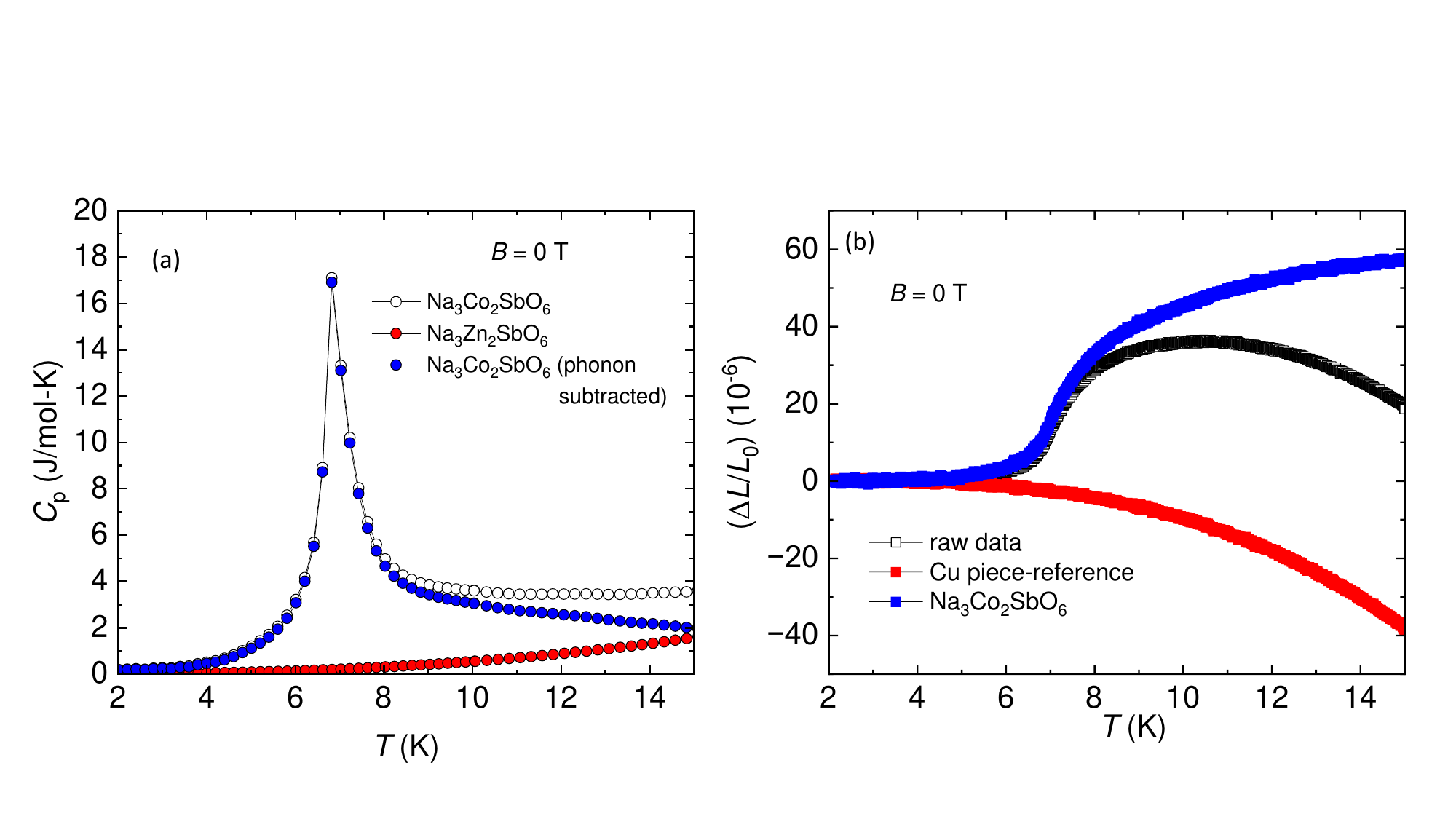}
	\caption{(a) $C_p$ vs. $T$ under zero magnetic field. The open black cirles are the raw data. To estimate the magnetic contribution (blue solid circles), we subtracted the phonon contributions taking Na$_3$Zn$_2$SbO$_6$, as the non-magnetic reference (red blue circles). (b) Given the small dimension of the crystal, it is necessary to subtract the background contribution from the temperature dependent dilatometry curves. To do so, we prepared a rectangular Cu piece of similar dimension as of the Na$_3$Co$_2$SbO$_6$ crystal and subtracted the background contributions.  
		\label{FigS4}}
\end{figure}

\begin{table}[htb!]
	\centering
	\caption{Zero-field length changes related to the AFM ordering in various honeycomb materials}
	\begin{tabular}{|c|c|}
		\hline
		\textbf{Compound} & \textbf{Relative length change in zero magnetic field} \\
		\hline
		BaCo$_2$(AsO$_4$)$_2$ & 4$\times$ 10$^{-4}$~\cite{MukharjeeL140407} \\
		\hline
		$\alpha$-RuCl$_3$ & 1$\times$ 10$^{-4}$~\cite{Gass2020} \\
		\hline
		BaCo$_2$(PO$_4$)$_2$ & 0.8$\times$ 10$^{-4}$~\cite{Wang2023} \\
		\hline
		Na$_2$Co$_2$TeO$_6$ & 0.2$\times$ 10$^{-4}$~{\cite{ArnethL140402}} \\
		\hline
		Na$_3$Co$_2$SbO$_6$ & 0.6$\times$ 10$^{-4}$~[This Work] \\
		\hline
	\end{tabular}
	\label{Tab1}
\end{table}%

\begin{figure*}
	\centering
	\includegraphics[width = 0.8\textwidth]{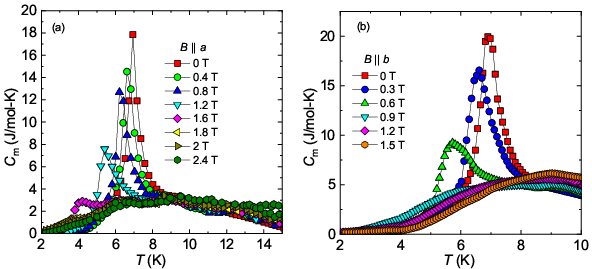}
	\caption{$C_m$ vs. $T$ at different magnetic fields for (a) $B \parallel a$ and (b) $B \parallel b$ orientations.}
	\label{FigS5}
\end{figure*}

\begin{figure*}[p]
	\centering
	\includegraphics[width = 0.8\textwidth]{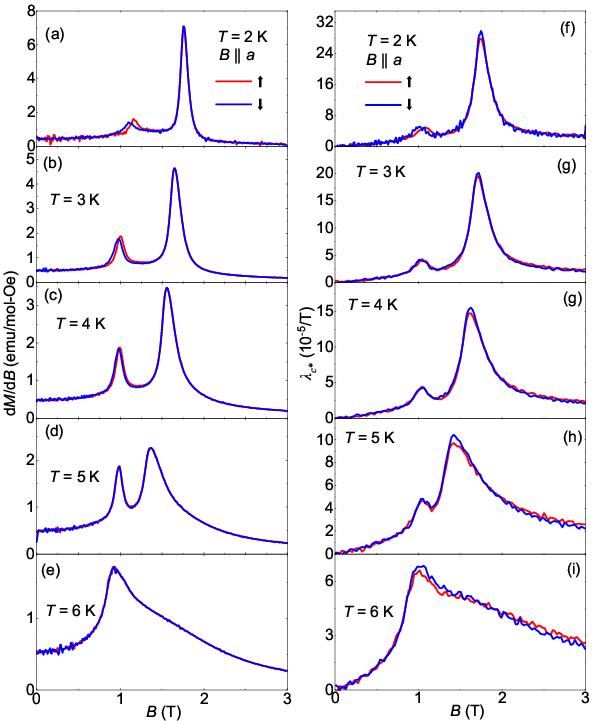}
	\caption{(a)-(e) Field variation of the differential susceptibility  and (f)-(i) $\lambda_{c^*}$ during increasing and decreasing field sweeps for $B \parallel a$.}
	\label{FigS6}
\end{figure*}

\clearpage

\begin{figure}
	\centering
	\includegraphics[width= 0.8\textwidth]{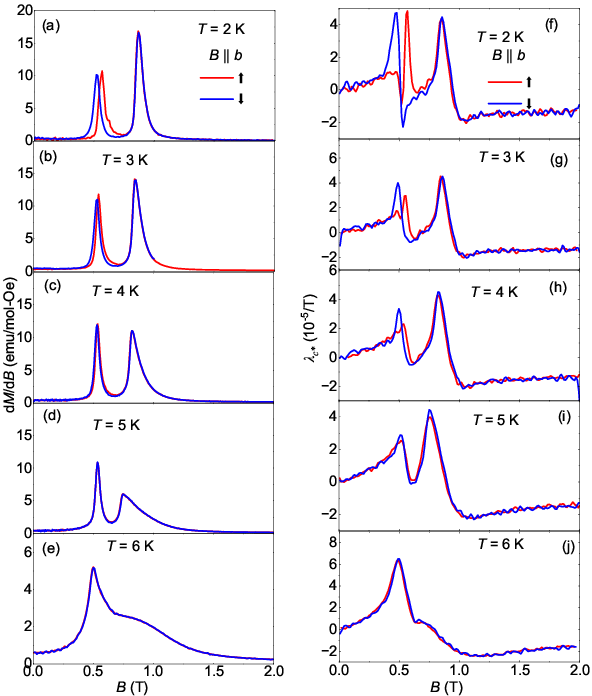}
	\caption{(a)-(e) Field variation of the differential susceptibility  and (f)-(i) $\lambda_{c^*}$ during increasing and decreasing field sweeps for $B \parallel b$.
		\label{FigS7}}
\end{figure}

\begin{figure}
	\includegraphics[width= 12 cm]{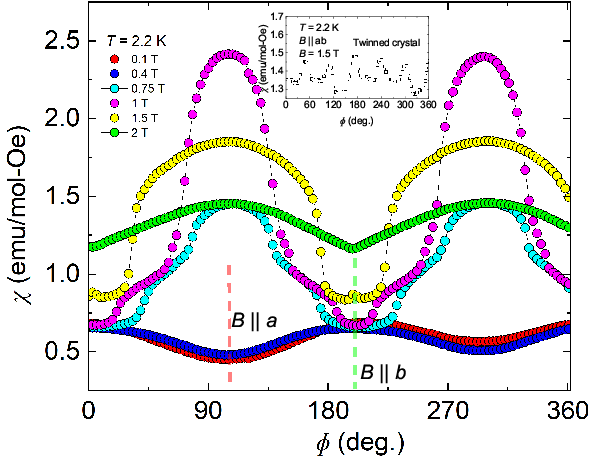}
	\caption{Angle-dependent magnetic susceptibility measured at $T = 2.2$~K, while the crystal is rotated in the $ab$-plane in different magnetic fields showing two-fold symmetry for 0\,T $\leq B \leq$ 2\,T. The dashed red and green lines indicate the positions of crystals for the two high symmetry directions. Inset: Same measurements conducted on a twinned crystals which clearly deviates from the $C_2$-rotation, and rather shows a six-fold symmetry. 
		\label{FigS8}}
\end{figure}

\begin{figure}[p]
	\includegraphics[width= \linewidth]{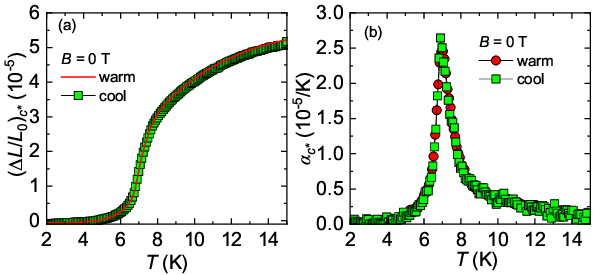}
	\caption{Warming and cooling curves under zero magnetic field for the relative length change (a) and its corresponding linear coefficient of thermal expansion (b) show no difference which supports the continuos nature of this transition.
		\label{FigS10}}
\end{figure}
\vspace{1em}
\end{document}